\documentclass[showpacs,prb,twocolumn]{revtex4}

\usepackage{amsfonts}
\usepackage{graphicx}
\usepackage{amsmath}
\usepackage{amssymb}
\usepackage{latexsym}
\usepackage{psfrag}

\begin{document}
\title{Origin of conductance quantization in disordered graphene ribbons}
\author{S. Ihnatsenka}
\affiliation{Department of Physics, Simon Fraser University, Burnaby, British Columbia, Canada V5A 1S6}
\author{G. Kirczenow}
\affiliation{Department of Physics, Simon Fraser University, Burnaby, British Columbia, Canada V5A 1S6}

\begin{abstract}
We present numerical studies of conduction in graphene nanoribbons with different types of disorder. We find that even when defect scattering depresses the conductance to values two orders of magnitude lower than $2e^2/h$, equally spaced conductance plateaus occur at moderately low temperatures due to enhanced electron backscattering near subband edge energies if bulk vacancies are present in the ribbon. This work accounts quantitatively for the surprising conductance quantization observed by  Lin et al. [Phys. Rev. B 78, 161409 (2008)] in ribbons with such low conductances.
\end{abstract}

\maketitle
   
Graphene nanoribbons (NR's) are attracting much experimental\cite{Lin08, Han07, Koskinen09} and theoretical\cite{Dresselhaus96, Onipko08, Son06, Evaldsson08, Mucciolo09,  Areshkin07, Yamamoto08, Igor08} interest due to their unique properties stemming from the linear, massless Dirac-like spectrum of the underlying honeycomb lattice. In common with other quasi one-dimensional (1D) ballistic nanostructures, {\it ideal} NR's are expected to exhibit conductances quantized in integer multiples of the conductance quantum $2e^2/h$ due to electron transmission via subbands that arise from  lateral confinement of electronic states in the NR. Recently Lin \textsl{et al.}\cite{Lin08} reported the first experimental observation of conductance quantization in NR's. However, surprisingly, the conductance steps that they observed were {\em orders of magnitude smaller} than $2e^2/h$. They decreased in height with increasing NR length and were present only at {\em moderately} low temperatures 10K $\lesssim$ T  $\lesssim$ 80K. Lin \textsl{et al.}\cite{Lin08} suggested that the quantized conductance steps that they observed may be due to different numbers of subbands in their NR's becoming populated with electrons as the back gate voltage in their system was varied. They attributed the low values of the quantized conductances ($<< 2e^2/h$) that they observed and their dependence on the length of the NR to low electron transmission probabilities through their device due primarily to scattering by defects. However, the quantum transport calculations reported to date\cite{Son06, Mucciolo09, Areshkin07, Yamamoto08} found conductance quantization to be {\em destroyed} by disorder even for NR's with much higher conductances (i.e., {\em much less disorder}) than those of the NR's studied by Lin \textsl{et al.}\cite{Lin08} Still it should be noted that the theoretical work was for narrower NR's than those studied by Lin \textsl{et al.}\cite{Lin08}  and transport should be more sensitive to disorder in narrower NR's. Furthermore the possibility of quantized conductances in strongly disordered NR's has not been the subject of systematic theoretical investigations. Thus the origin of the quantized conductances observed by Lin \textsl{et al.}\cite{Lin08} has remained an open question.

The purpose of this paper is to investigate theoretically how different scattering mechanisms affect electron transport in {\em wide} disordered NR's such as those of Lin \textsl{et al.}\cite{Lin08} and to clarify under which conditions quantized conductances {\em much smaller} than $2e^2/h$ can occur in such systems. We find that electron scattering by C atom vacancies at moderately low temperatures (in the presence of edge disorder) can account quantitatively for the quantized conductances observed by  Lin \textsl{et al.}\cite{Lin08} The underlying mechanism that we identify is modulation of the NR conductance by enhanced electron back-scattering by vacancies whenever a subband edge crosses the Fermi level.    

We concentrate on three disorder types, namely bulk vacancies, edge imperfections and long-range potentials due to charged impurities. Other disorder types such as weak short-range potentials due to neutral impurities,\cite{Endo01} and lattice distortions\cite{Son06, Koskinen09}, may be present, but the three disorder cases to be discussed here have the strongest impact on transport through NR's and thus are more relevant to the strong conductance suppression reported in Ref. \onlinecite{Lin08}. Among the three, only bulk vacancies scale the heights of different conductance steps uniformly. Thus their presence appears crucial for the observation of conductance quantization in NR's with strong disorder. 


 We describe NR's by the standard tight-binding Hamiltonian on a honeycomb lattice,
\begin{equation}
 H=\sum_{i}\epsilon _{i}a_{i}^{\dag }a_{i}-\sum_{\left\langle i,j\right\rangle }t_{ij}\left( a_{i}^{\dag }a_{j}+h.c. \right),
 \label{eq:hamiltonian}
\end{equation}%
where $\epsilon _{i}$ is the on-site energy and $t_{ij}=t=2.7$ eV is the matrix element between nearest-neighbor atoms. This Hamiltonian is known to describe the $\pi$ band dispersion of graphene well at low energies.\cite{Reich02} Spin and electron interaction effects are outside of the scope of our study. 
Bulk vacancies and edge disorder are introduced by randomly removing carbon atoms and setting appropriate hopping elements $t_{ij}$ to zero. It is assumed that atoms at the edges are always attached to two other carbon atoms and passivated by a neutral chemical ligand, such as hydrogen. The bulk and edge disorder are characterized by the probability of the carbon atoms being removed, $p^b$ and $p^e$, respectively. $p^b$ is normalized relative to the whole sample, while $p^e$ is defined relative to an edge only. The long-range potential due to charged impurities is approximated by a Gaussian form\cite{Mucciolo09, Yamamoto08} of range $d$: 
$\epsilon_i=\sum_{r_0} V_0 \text{exp}({-\left|r_i-r_0\right|^2}/{d^2})$,
where both the amplitude $V_0$ and coordinate $r_0$ are generated randomly.  

\begin{figure*}[t]
\includegraphics[scale=0.9]{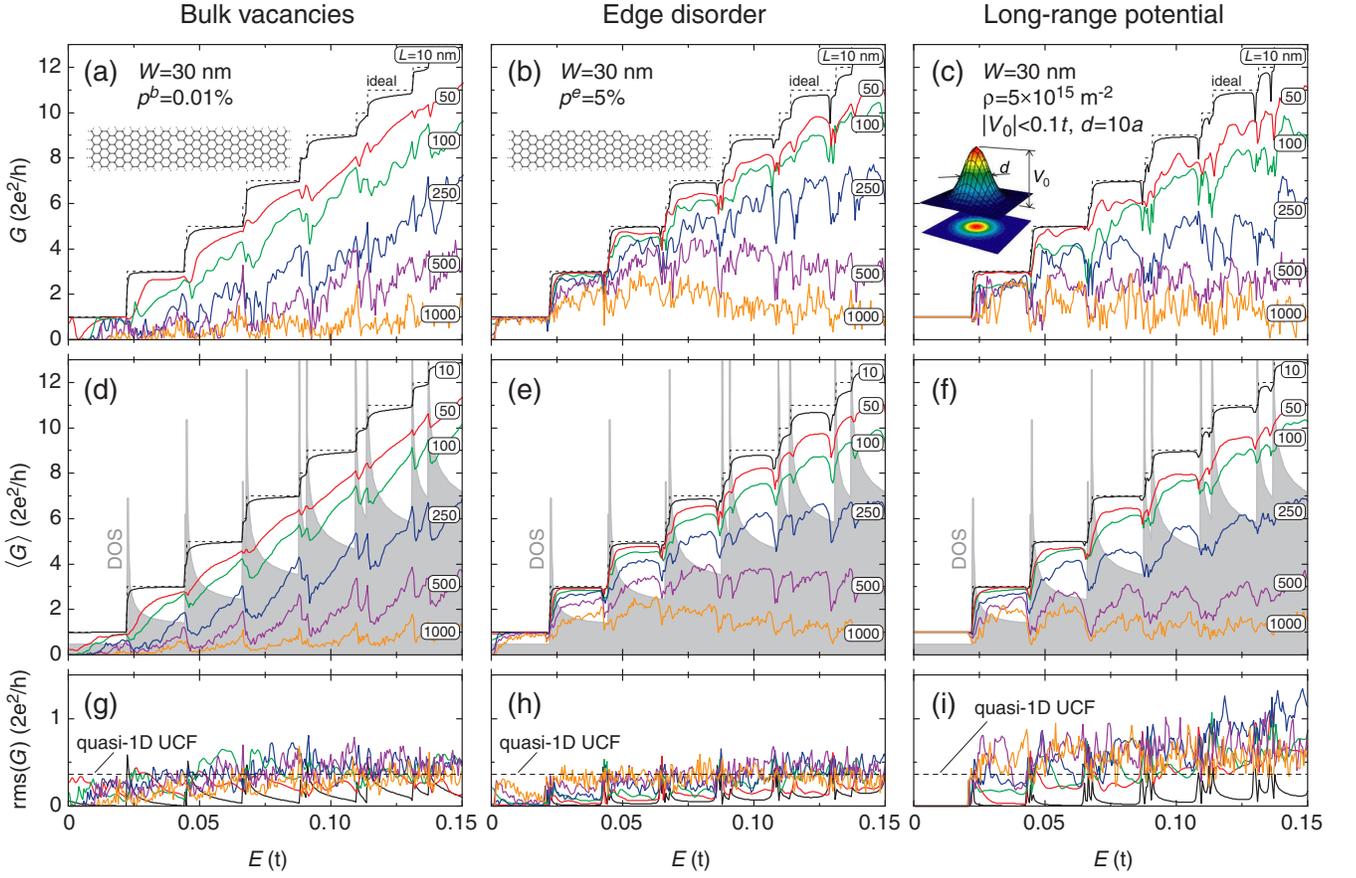}
\caption{(color online) Conductance (a)-(c), average conductance (d)-(f) and conductance fluctuations (g)-(i) as a function of energy for the graphene ribbon of width $W=30$ nm and different lengths $L=10...1000$ nm. Left panel corresponds to bulk vacancies, middle panel is for edge disorder while right one shows the effect of long-range potential. Parameters of disorder with representative illustrations are given in the plots (a)-(c). The dotted lines in (a)-(f) show the conductance quantization for the ideal ribbon. The gray filled areas in (d)-(f) denote $DOS$ for the ideal ribbon. The dashed lines in (g)-(i) mark the universal value of the conductance fluctuations for quasi-1D systems.\cite{Lee85} Temperature $T=0$. $t=2.7$ eV; $a=0.142$ nm.}
\label{fig:1}
\end{figure*}

 In the linear response regime the conductance of the NR is given by the Landauer formula\cite{Landauer}
\begin{equation}
	G = -\frac{2e^{2}}{h} \int_{0}^{\infty }dE^{\prime} \; T(E^{\prime}) \frac{\partial f_{FD}(E^{\prime}-E)}{\partial E^{\prime}}.
	\label{eq:conductance}
\end{equation}
$T(E)$ is the total transmission coefficient and $f_{FD}(E)$ is the Fermi-Dirac function. $T(E)$ is calculated by the recursive Green's function method, see Ref. \onlinecite{Igor08} for details. Fluctuations of the conductance are defined by  $rms(G) = ( \left\langle G^2\right\rangle - \left\langle G\right\rangle^2 )^{1/2}$, where $\left\langle \right\rangle$ denotes averaging over an ensemble of samples with different realizations of disorder. For the results presented below, averaging was carried out over ten realization for each disorder type.

 To investigate the transport properties of disordered NR's we choose geometries similar to ones studied experimentally.\cite{Lin08} The disorder is assumed to exist in a finite ribbon of width $W$ and length $L$. This ribbon is attached at its two ends to semiinfinite leads represented by ideal NR's of width $W$. The edge (host) configuration is taken as armchair in the following. Representative disorder geometries are shown in the insets in Figs. \ref{fig:1}(a)-(c). 

Fig. \ref{fig:1} shows the effect of different disorder types on conduction in NR's. For each disorder type we keep the defect concentration and strength and the ribbon width fixed ($W=30$ nm) and vary its length $L$. As $L$ increases, and the number of scattering centers grows, the conductance decays and quantization steps are destroyed. 

For bulk vacancy disorder, even a small concentration of the defects affects the conductance strongly; see Figs. \ref{fig:1}(a),(d). Apart from reduced conductances, the disorder results in sample-specific conductance fluctuations, Figs. \ref{fig:1}(a),(g), whose amplitude is of order $e^2/h$, independent of energy or NR length. This is a quantum interference effect similar to the universal conductance fluctuations (UCF's) of mesoscopic metals.\cite{Lee85} The particular value of the conductance depends sensitively on the electron energy, ribbon length and locations of the vacancies. Since the vacancies are distributed over the whole sample, intra-subband scattering predominates. Thus the conductance in Figs. \ref{fig:1}(a),(d) (coarse grained in energy to smooth out UCF's) scales {\em uniformly} with NR length \textit{L}, i.e., in a similar way for all subbands. This resembles bulk island scattering in conventional quantum wires.\cite{Nikolic94} 

By contrast, for edge disorder in Figs. \ref{fig:1}(b),(e) the conductance scales {\em non}-uniformly: Defects at the boundaries scatter electrons equally into all subbands resulting in stronger suppression of the conductance at higher energies $E$ where more subbands are available; see e.g., $L=1000$ nm ribbon in Figs. \ref{fig:1}(b),(e).

Potential inhomogeneities due to charged impurities lead to the appearance of electron and hole puddles in NR's.\cite{Martin08} Scattering by the potential inhomogeneities results in subband mixing that smears conductance steps, Figs. \ref{fig:1}(c),(f). As the subband number increases, intervalley scattering becomes more effective with stronger backscattering of higher subband states in long ribbons. The first subband, however, is not affected by the long-range potential because of internal phase structures of its wave function that make the scattering amplitude vanish.\cite{Yamamoto08} The conductance fluctuations are roughly twice as strong as for bulk vacancy and edge disorder, Figs. \ref{fig:1}(g)-(i). This may be due to weaker inter-valley scattering for which particles at $K$ and $K^{\prime}$ Dirac points contribute independently to the UCFs.\cite{Dresselhaus96, Areshkin07} The fluctuation amplitudes agree reasonably well with the value for UCF's in quasi-1D systems\cite{Lee85}, $0.729\frac{e^2}{h}$.

A prominent effect of all disorder types is the formation of a conductance dip when the Fermi level crosses a subband edge. This is most obvious in the averaged conductance $\left\langle G\right\rangle$,  Figs. \ref{fig:1}(d)-(f). The origin is the strong intersubband scattering caused by defects, where an electron in a state $\left|nk\right\rangle$ scatters into another state $\left|n^{\prime}k^{\prime}\right\rangle$. It can be understood physically by considering the Fermi Golden rule expression for the scattering time $\tau$\cite{Davies_book}:
\begin{equation}
	\frac{1}{\tau}=\frac{2\pi}{\hbar}\sum_{n^{\prime}} \left|\left\langle nk \left|H^{\prime}\right| n^{\prime}k^{\prime} \right\rangle\right|^2 \rho_{n^{\prime}}\left( E \right).
	\label{eq:goldenrule}
\end{equation}
Here $H^{\prime}$ is the perturbation due to defects and $\rho_{n^{\prime}}\left( E \right)$ the density of states of the $n^{\prime}$-th subband. Assuming that $\left|\left\langle nk \left|H^{\prime}\right| n^{\prime}k^{\prime} \right\rangle\right|^2$ is independent of the band index $n^{\prime}$, the scattering rate $1/\tau$ is seen to be proportional to the total density of states of the ribbon, $\rho(E)=\sum_{n^{\prime}} \rho_{n^{\prime}}(E)$. For a perfect ribbon the dispersion relation can be approximated by a parabolic function if $k$ is small and $\left|n\right|>1$.\cite{Onipko08} Therefore, $\rho(E)$ diverges at subband thresholds $E_{n^{\prime}}$ as $(E-E_{n^{\prime}})^{1/2}$. This agrees with the numerically calculated density of states for the tight-binding Hamiltonian \eqref{eq:hamiltonian}, see the gray areas in Figs. \ref{fig:1}(d)-(f). Thus, the scattering time $\tau$ is strongly reduced when the Fermi energy approaches a subband threshold $E_{n^{\prime}}$ and the transmission of electrons in the $n$-th subband is strongly suppressed due the scattering  into the other $n^{\prime}$ subbands. As a result, the conductance shows dips at the subband edges.

\begin{figure}[t]
\includegraphics[scale=1.0]{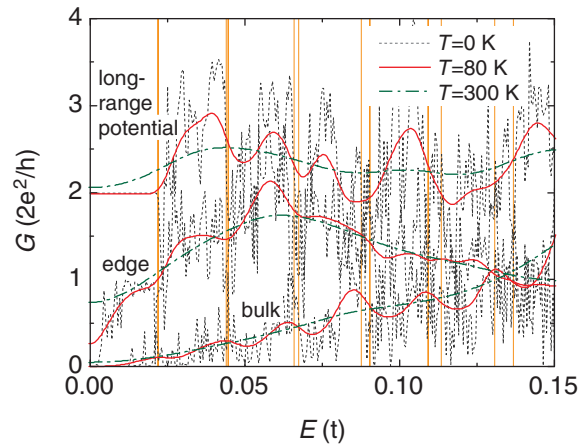}
\caption{(color online) Conductance through disordered ribbons as a function of the Fermi energy for temperatures $T=0, 80, 300$ K. The ribbons have width $W=30$ nm and length $L=1000$ nm; parameters of disorder are listed in Fig. \ref{fig:1}. Thin solid vertical lines correspond to energies when number of subbands changes by one. For the sake of clarity the curves for long-range potential are shifted upward by $2e^2/h$.}
\label{fig:2}
\end{figure}

As the temperature increases, the conductance fluctuations are smeared out and the dips in the conductance associated with enhanced electron back scattering when the Fermi level crosses subband edges become clearly visible for temperatures $T$  not greatly exceeding the subband energy separation, $ 4\pi k_B T \approx \Delta E = E_{n+1}-E_n$. For graphene ribbons 30 nm wide $\Delta E \approx 0.02t=54$ meV, see Fig. \ref{fig:1}, that corresponds to $T \approx \frac{\Delta E}{4\pi k_B}=50$ K. Above this temperature the conductance dips become gradually smeared, but well below it the conductance may be dominated by UCF's of the disordered ribbon. This estimate is in good agreement with calculations presented in Fig. \ref{fig:2}, where ribbons with different disorder types are subjected to $T=0, 80, 300$ K: The conductance dips at subband edges manifest as the smooth conductance oscillations that are clearly visible for $T=80$ K. They are very regular and are superimposed on a smoothly rising background for the case of the bulk vacancies, but appear very distorted when the edge disorder or long-range potential introduced, except for the first two or three oscillations for the case of the edge disorder. Since the effects of long-range disorder are similar to those of edge disorder (see Figs. \ref{fig:1}(e),(f)) we shall not consider the long-range disorder further here.

The conductances of the NR's measured by Lin \textsl{et al.}\cite{Lin08} were 65-260 times smaller than the conductance quantum and also much smaller than the conductances of the model systems studied above in  Figs. \ref{fig:1} and \ref{fig:2}. However, the ideas developed above apply equally well to the lower conductance regime in which the experiments were carried out and are able to account quantitatively for the conductance quantization that Lin \textsl{et al.}\cite{Lin08} observed. We demonstrate this next by presenting simulations for NR's with the same sizes as in the experiments\cite{Lin08} and with defect concentrations chosen to yield low conductances similar to those measured by Lin et al.\cite{Lin08}

\begin{figure}[t]
\includegraphics[scale=1.0]{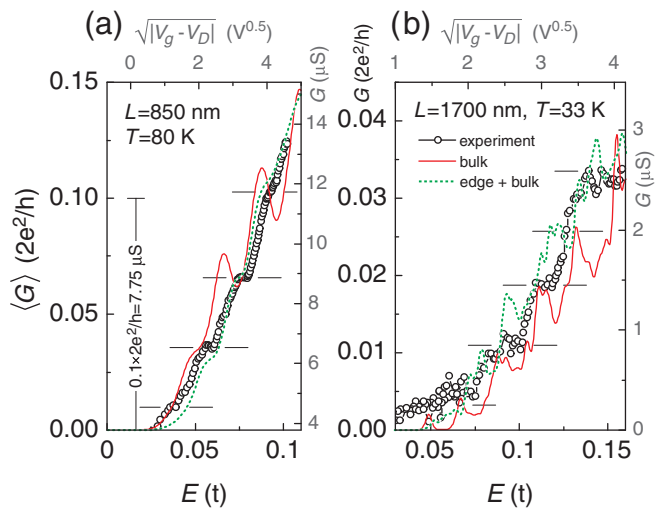}
\caption{(color online) Comparison of theoretical and experimental data for the ribbons of $W=30$ nm and $L=850, 1700$ nm. Experimental data is adopted from Ref. \onlinecite{Lin08}. Theoretical calculations are performed for two disorder cases: (a) - only bulk vacancies $p^b=4\times 10^{-4}$ (solid red curve), and combination of edge $p^e=2$ and bulk $p^b=10^{-4}$ disorder (dotted green curve); (b) - $p^b=8\times 10^{-4}$, and combination of $p^e=3.5$ and $p^b=2\times 10^{-4}$. The gate voltage $V_g$ is scaled to produce better fit; note that $E \sim \sqrt{V_g}$ as discussed in Ref.\onlinecite {Li08}.}
\label{fig:3}
\end{figure}

Fig. \ref{fig:3} shows the calculated conductances $\left\langle G\right\rangle$ of the disordered ribbons along with the experimental data from Ref. \onlinecite{Lin08}. The features in the theoretical plots that match the experimental conductance plateaus are the conductance dips that are due to enhanced electron back scattering at the energies of the subband edges of the nanoribbon that we have already discussed in connection with Figs.  \ref{fig:1} and \ref{fig:2}.  The agreement between theory and experiment is remarkable especially for the heights of the conductance plateaus. From the theoretical point of view, there are several detailed scenarios that might result in this behavior. Unfortunately, it is not possible to rule out any of them because the experiment gives no information regarding which disorder is type actually realized. Therefore, we propose that the dominant scattering mechanism might be either due to bulk vacancies alone or a combination of rough edges with a lower concentration of bulk vacancies. The presence of the latter is crucial because they equalize the differences between the conductances of the different plateaus making them equidistant. In particular, we found that $p^b=4\times10^{-4}$ bulk vacancies are enough to reduce the conductances of the quantized plateaus by a factor of 65 relative to the conductance quantum $\frac{2e^2}{h}$, in accord with the experiment\cite{Lin08}, see solid red line in Fig. \ref{fig:3}(a). This means that one in 2500 carbon atoms is removed, which seems plausible. The other scenario consists of distorted edges with two rows of carbon atoms removed on average along the boundaries and also one in 10000 bulk carbons removed, $p^e=2$ and $p^e=10^{-4}$, see dashed green line in Fig. \ref{fig:3}(a). For the longer $L=1700$ nm ribbon the height of conductance steps drops to a factor 260 lower than the conductance quantum, Fig. \ref{fig:3}(b). This implies defect concentrations twice those of the shorter $L=850$ nm experimental ribbon. The lower temperature in Fig. \ref{fig:3}(b) results in stronger conductance fluctuations than in Fig. \ref{fig:3}(a); the fourth plateau being not discernible in the experimental data\cite{Lin08}  in Fig. \ref{fig:3}(b), may also be due in part to a particular disorder configuration.  However, all visible conductance plateaus are due to subband formation associated with particle motion quantized in the transverse direction. At much lower temperatures in our simulations these conductance plateaus are not discernible due to UCF's and they also disappear completely at room temperature, behavior similar to that in Fig. \ref{fig:2}, and completely consistent with the data of Lin \textit{et al.}\cite{Lin08}.

In conclusion, our quantum transport calculations have shown that equally spaced quantized conductance plateaus should be observable in disordered graphene nanoribbons even for conductance values much smaller than the conductance quantum $2e^2/h$ at temperatures comparable to subband energy spacings. The plateaus are due to enhanced electron back scattering by defects  at energies near subband edges. Deviations from equal spacing of the conductance plateaus can occur depending on the defect configurations in particular experimental samples. These findings provide a microscopic explanation of the conductance quantization of  graphene nanoribbons observed by Lin \textit{et al.}\cite{Lin08} and suggest that the observed conductance quantization\cite{Lin08} can be regarded as a signature of subband formation.

This work was supported by NSERC, CIFAR and WestGrid. 
We thank I. V. Zozoulenko for discussions.

\end{document}